\documentclass{PoS}
\input{babarsym}
\usepackage{amsmath} 

\title{Search for a low-mass Higgs boson ($A^0$) at \babar\/}

\ShortTitle{Search for a low-mass Higgs boson ($A^0$) at \babar\/}

\author{\speaker{Arafat Gabareen Mokhtar}\thanks{On behalf of the \babar\ Collaboration}\\
        SLAC National Accelerator Laboratory\\
	2575 Sand Hill Rd, Menlo Park, CA 94025, USA\\
        E-mail: \email{mokhtar@slac.stanford.edu}}

\abstract{The \babar\ Collaboration has performed three searches for a
light Higgs boson, $A^0$, in radiative Upsilon ($\Upsilon$) decays:
$\Upsilon(3S)\to\gamma A^0$, $A^0\to\tau^+\tau^-$;
$\Upsilon(nS)\to\gamma A^0$, $A^0\to\mu^+\mu^-$ ($n=2,3$); and
$\Upsilon(3S)\to\gamma A^0$, $A^0\to$ invisible. Such a Higgs boson
($A^0$) appears in the Next-to-Minimal Supersymmetric extensions of
the Standard Model, where a light $CP$-odd Higgs boson couples
strongly to $b$-quarks. The searches are based on data samples that
consist of $122\times 10^6$ $\Upsilon(3S)$ and $99\times 10^6$
$\Upsilon(2S)$ decays, collected by the \babar\ detector at the SLAC
National Accelerator Laboratory. The searches reveal no evidence for
an $A^0$, and product of branching fractions upper limits, at $90\%$
C.L., of $(1.5-16)\times 10^{-5}$, $(0.44-44)\times 10^{-6}$, and
$(0.7-31)\times 10^{-6}$ were obtained for these searches,
respectively. Also, we set the upper limits ${\mathcal
{B}}(\eta_b\to\tau^+\tau^-)<8\%$ and ${\mathcal
{B}}(\eta_b\to\mu^+\mu^-)<0.9\%$.}

\FullConference{The 2009 Europhysics Conference on High Energy Physics,\\
		 July 16 - 22 2009\\
		 Krakow, Poland}

\begin{document}

\section{Introduction}
The origin of mass in the Standard Model (SM) arises through the Higgs
mechanism where an elementary scalar acquires a vacuum expectation
value. The minimal SM has a single Higgs doublet model; however, in
many theories beyond the SM there may be more elaborate Higgs sector.
The minimal supersymmetric Standard Model (MSSM) requires two Higgs
doublets to generate masses for all SM fermions~\cite{Haber:1984rc}
and in the next-to-minimal supersymmetric Standard Model (NMSSM),
there is one additional complex singlet scalar in addition to the two
Higgs doublets~\cite{ref:dgm}. The NMSSM naturally addresses several
over the problems of the MSSM such as the origin of the $\mu$ and
$B_\mu$ terms and is considered one of the prime extensions to the
Standard Model. Frequently in the NMSSM there is a very light
pseudo-scalar state, $A^0$, with a mass less than 10 \gev\ and this
state can couple weakly to the SM fermions with a coupling
proportional to the mass of the fermion. This motivates a search for
the decay of $\Upsilon \rightarrow \gamma A^0$ which occurs whenever
the mass of the $A^0$ is less than the $b\bar{b}$ production
threshold~\cite{ref:wilcek}. This search is presented in this
proceeding at \babar\ with $122 \times 10^6$ $\Upsilon(3S)$ and $99
\times 10^6$ $\Upsilon (2S)$ states.

\section{{\bf \boldmath Search for $\Upsilon(3S)\to\gamma A^0$, $A^0\to\tau^+\tau^-$}}
We study the decays
$\Upsilon(3S)\to\gamma\tau^+\tau^-$~\cite{Aubert:2009cka} using
$122\times 10^6$ $\Upsilon(3S)$ decays. In this analysis the search
for $A^0$ is extended for a wider mass range with respect to a
previous CLEO Collaboration analysis~\cite{ref:cleo}. We scan for
peaks in the distribution of the photon energy, $E_\gamma$,
corresponding to peaks in the $\tau\tau$ invariant mass
$m_{\tau\tau}^2=m_{3S}^2-2m_{3S}E_\gamma$, where $m_{3S}$ is the
$\Upsilon(3S)$ mass and $E_\gamma$ is measured in the $\Upsilon(3S)$
rest frame. We quote branching fractions in the region
$4.03<m_{\tau\tau}<10.10$ \gevcc\/, but we exclude the region
$9.52<m_{\tau\tau}<9.61$ \gevcc\/, which corresponds to the expected
irreducible background of photons in the decay chain
$\Upsilon(3S)\to\gamma\chi_{bJ}(2P)$,
$\chi_{bJ}(2P)\to\gamma\Upsilon(1S)$, where $J=0,1,2$.

\begin{figure}[!htbp]
  \begin{center}
    \includegraphics[width=10.cm]{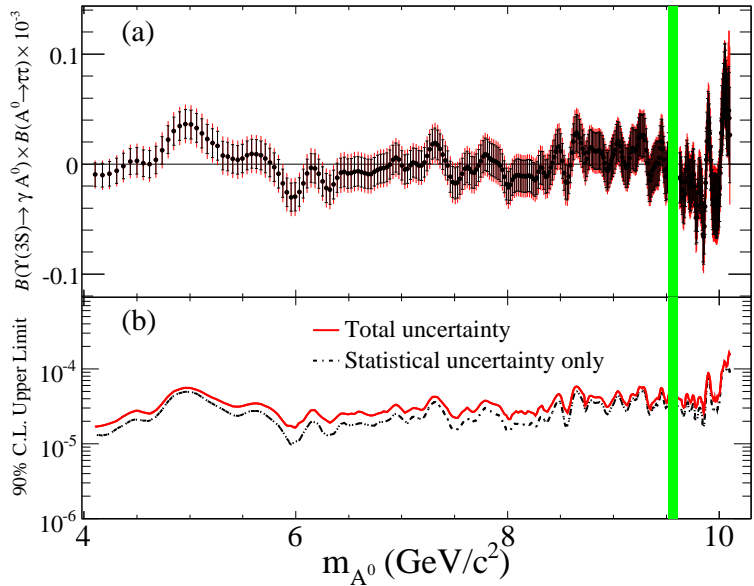}
    \caption{(a) The product of branching fractions
    (${\mathcal{B}}(\Upsilon(3S)\to\gamma A^0)\times
    {\mathcal{B}}(A^0\to\tau^+\tau^-)$) as a function of
    $m_{A^0}$. For each point, both the statistical uncertainty and
    the total uncertainty are shown. In (b), the corresponding 90$\%$
    C.L. upper limits on the product of the branching fractions versus
    the Higgs mass values are shown, with total uncertainty (solid
    line) and statistical uncertainty only (dashed line). The shaded
    vertical region represents the excluded mass range corresponding
    to the $\chi_{bJ}(2P)\rightarrow\gamma\Upsilon(1S)$ states.}
    \label{fig:1}
\end{center}
\end{figure}

The product branching fractions are shown in
Fig.~\ref{fig:1}(a). These results show no evidence for a narrow
resonance in the mass range under study. Bayesian upper limits on the
product of branching fractions, computed with a uniform prior at
$90\%$ C.L., are shown in Fig.~\ref{fig:1}(b). The upper limits on the
product branching fraction ${\cal {B}}(\Upsilon(3S)\rightarrow\gamma
A^0)\times {\cal {B}}(A^0\rightarrow\tau^+\tau^-)$ vary between
$(1.5-16)\times 10^{-5}$ at $90\%$ C.L. We also set the upper limit
${\cal{B}}(\eta_b\to\tau\tau)<8\%$ at $m_{\tau\tau}=9.389$ \gevcc\
using the ${\cal{B}}(\Upsilon(3S)\to\gamma\eta_b)$ from
Ref.~\cite{:2008vj}.

\section{Search for $\Upsilon(2S,3S)\to\gamma A^0$, $A^0\to\mu^+\mu^-$}
\begin{figure}[!htbp]
  \begin{center}
    \includegraphics[width=10.cm]{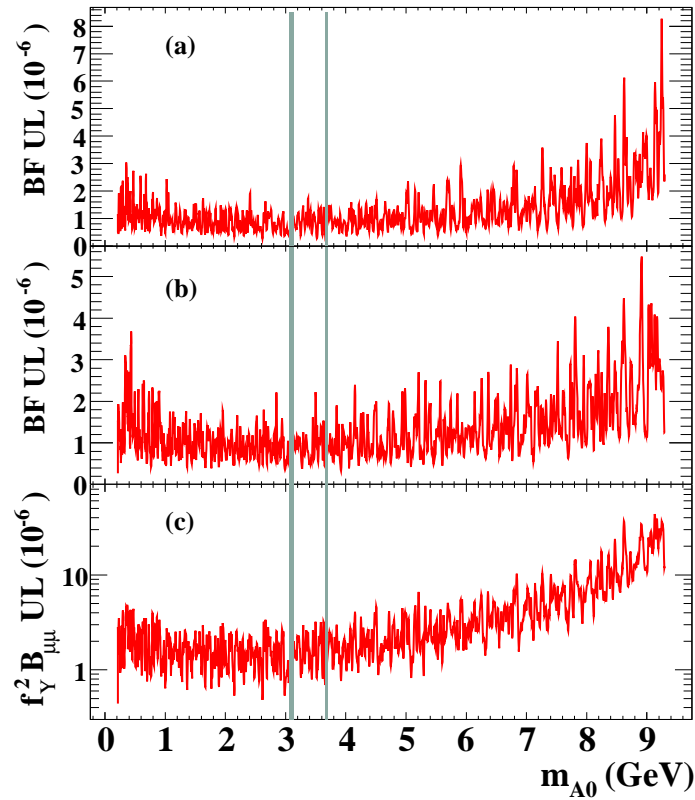}
    \caption{$90\%$ C.L. upper limits on (a)
    ${\mathcal{B}}(\Upsilon(2S)\to\gamma A^0)\times
    {\mathcal{B}}_{\mu\mu}$ where
    $\mathcal{B}_{\mu\mu}\equiv\mathcal{B}(A^0\to \mu\mu)$, (b)
    ${\mathcal{B}}(\Upsilon(3S)\to\gamma A^0)\times
    {\mathcal{B}}_{\mu\mu}$, and (c) effective coupling
    $f^2_\Upsilon\times {\mathcal{B}}_{\mu\mu}$ as a function of
    $m_{A^0}$. The shaded areas represent the $J/\psi$ and $\psi(2S)$
    regions, excluded from this search.}
    \label{fig:2}
\end{center}
\end{figure}

We search for $A^0$ in the decays $\Upsilon(nS)\to\gamma A^0$,
$A^0\to\mu^+\mu^-$ ($n=2,3$) using the \babar\ data sample which
contains $99\times 10^6$ $\Upsilon(2S)$ and $122\times 10^6$
$\Upsilon(3S)$ decays~\cite{Aubert:2009cp}. No significant excess of
events above the background in the range $0.212<m_{\mu\mu}<9.3$
\gevcc\/ was observed. The $90\%$ C.L. upper limits on the product
branching fractions as a function of $m_{A^0}$ for $\Upsilon(2S)$ and
$\Upsilon(3S)$ are shown in Fig.~\ref{fig:2}(a) and
Fig.~\ref{fig:2}(b), respectively, and span the range $(0.44-44)\times
10^{-6}$. We also compute the upper limit
${\cal{B}}(\eta_b\to\mu\mu)<0.9\%$ at $90\%$ C.L. The branching
fractions ${\mathcal{B}}(\Upsilon(nS)\to \gamma A^0)$ are related to
the effective coupling $f_{\Upsilon}$ of the bound $b$ quark to the
$A^0$~\cite{ref:wilcek,Mangano:2007gi,Nason:1986tr}. The effective
coupling $f^2_\Upsilon\times {\mathcal{B}}_{\mu\mu}$ as a function of
the $m_{A^0}$ is shown in Fig.~\ref{fig:2}(c).

\section{Search for $\Upsilon(3S)\to\gamma A^0$, $A^0\to\mathrm{invisible}$}
\begin{figure}[!htbp]
  \begin{center}
    \includegraphics[width=10.cm]{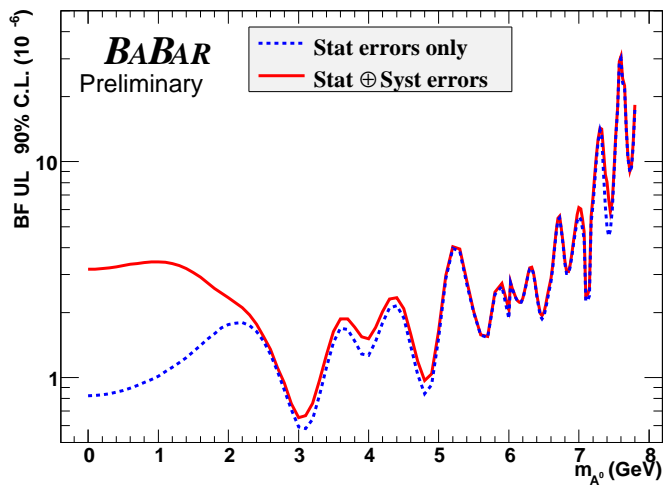}
    \caption{$90\%$ C.L. upper limits on
    ${\mathcal{B}}(\Upsilon(3S)\to\gamma A^0)\times
    {\mathcal{B}}(A^0\to\mathrm{invisible})$ as a function of
    $m_{A^0}$. The limits are shown with the statistical uncertainties
    only (dashed line) and the total uncertainties (solid line). }
    \label{fig:3}
\end{center}
\end{figure}

We search for $A^0$ produced in single-photon decays of the
$\Upsilon(3S)$ resonance through the process $\Upsilon(3S)\to\gamma
A^0$, $A^0\to\mathrm{invisible}$~\cite{BAD2073}. The analysis is based
on a \babar\ dataset consisting of $122\times 10^6$ $\Upsilon(3S)$
decays. We search for events with a high-energy photon and no other
particles that are consistent with a two-body decay of the
$\Upsilon(3S)$. We find no evidence for such a process and set $90\%$
C.L. upper limits on the branching fraction
$\mathcal{B}(\Upsilon(3S)\to\gamma A^0)\times
{\mathcal{B}}(A^0\to\mathrm{invisible})$ at $(0.7-31)\times 10^{-6}$
in the mass range $m_{A^0}<7.8$ \gevcc\/, as shown in
Fig.~\ref{fig:3}.

\end{document}